\documentclass[prd,preprint]{revtex4}%
\usepackage{graphicx}
\usepackage{bm}
\usepackage{epsf}
\usepackage{url}
\usepackage{rotating}
\usepackage{epsfig,graphics,rotate,color}
\usepackage{wrapfig}
\usepackage{amssymb}
\usepackage{amsmath}
\usepackage{amsfonts}
\usepackage{array,hhline,dcolumn}%
\usepackage{multirow}
\usepackage{longtable}
\usepackage{lscape}
\usepackage{graphicx}
\usepackage{sidecap}
\providecommand{\U}[1]{\protect\rule{.1in}{.1in}}
\begin{document}

\title{The Desktop Muon Detector:  A simple, physics-motivated machine- and electronics-shop project for university students}

\author{S.N. Axani}
\email{saxani@mit.edu} % optional
\affiliation{Physics Department, Massachusetts Institute of Technology, Cambridge, MA 02139}

\author{J.M. Conrad}
\email{conrad@mit.edu} % optional
\affiliation{Physics Department, Massachusetts Institute of Technology, Cambridge, MA 02139}

\author{C. Kirby}
\email{ckirby@mit.edu} % optional
\affiliation{Physics Department, Massachusetts Institute of Technology, Cambridge, MA 02139}

\date{\today}

\begin{abstract}
This paper describes an undergraduate-level physics project that incorporates various aspects of machine- and electronics-shop technical development. The desktop muon detector is a self-contained apparatus that employs a plastic scintillator as the detection medium and a silicon photomultiplier for light collection. These detectors can be battery powered and used in conjunction with the provided software to make interesting physics measurements.  The total cost per detector is approximately \$100.

\end{abstract}

\maketitle

\section{Introduction}

\begin{figure}[h!]
 \centering
\includegraphics[width=0.8\columnwidth]{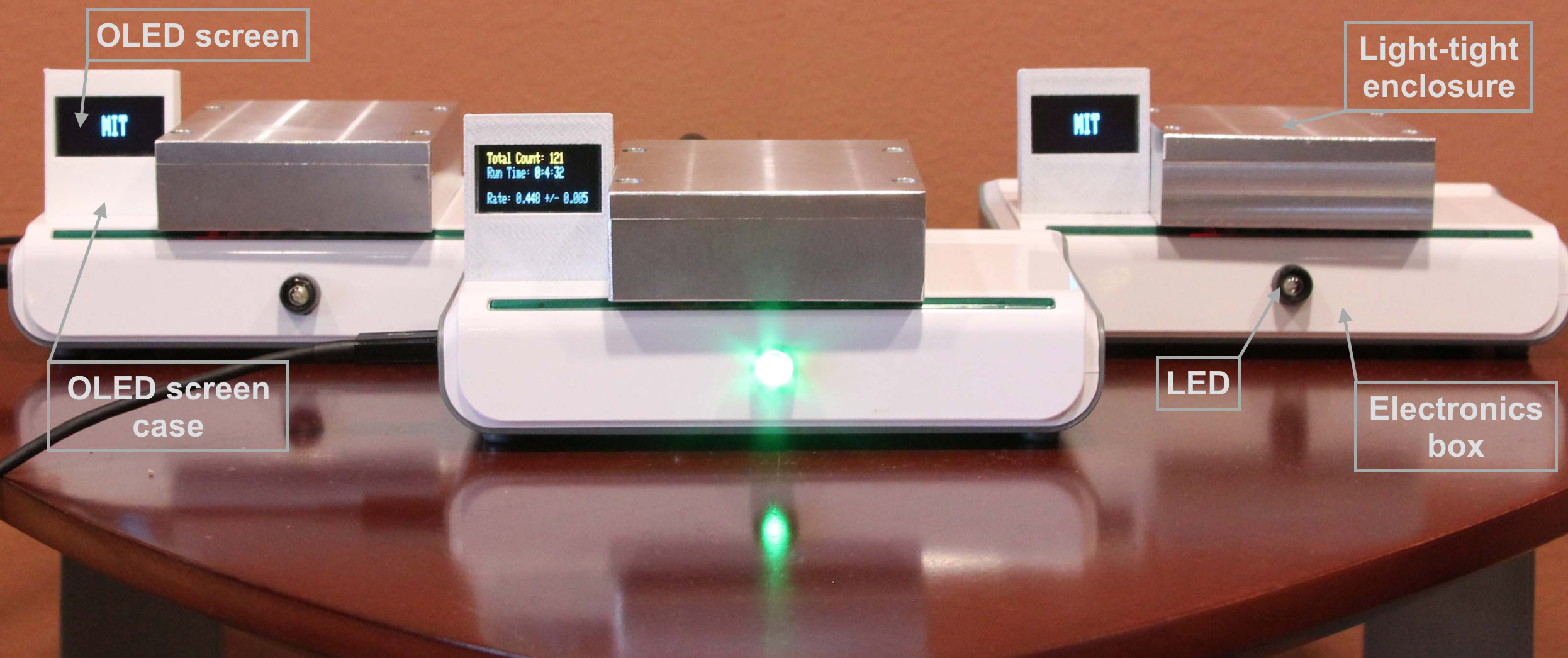}
  \caption{An array of desktop muon detectors and the corresponding components. When a muon passes through one of the light-tight aluminum boxes, the electronics box records the event and displays the information on the 0.96-inch organic light-emitting diode (OLED) screen. The green light-emitting diode (LED) in the front flashes for a period governed by the number of photons observed by the photodetector. The data can then be transmitted to a computer via a mini-USB cable.}
  \label{fig:array}
\end{figure}

This paper describes an undergraduate project to produce a compact, self-contained ``desktop muon counter.''    The muon detector is contained in a small light-tight enclosure that measures 2.75$\times$3.00$\times$1.00 in$^3$.  This sits on a small electronics box that performs the data acquisition and readout for the detector. The process of making the counters and the readout will teach students valuable skills in machining and in constructing and debugging electronics. We can use the end product as a single device (or in sets) to make interesting physics measurements or give introductory physics demonstrations. As an example, we present a measurement made at Fermi National Accelerator Laboratory (Fermilab).  In the supplementary material \cite{sup}, we provide the computer-aided design (CAD) drawings for machining, the files for the printed circuit boards (PCBs), the code required to program the microcontroller, and the Python program to write the detector data to a computer. The overall cost per counter is about \$100. A cost breakdown is also supplied in the supplementary material (Purchasing\_list.xml).  An array of desktop muon counters is shown in Fig.~\ref{fig:array}. 

This paper is inspired by the Phys. 063 class at Swarthmore College, ``Procedures in Experimental Physics,'' which was taken by one of the authors. This class introduces students to ``the techniques, materials, and the design of experimental apparatus; shop practice; printed circuit design and construction''  \cite{Swat}. This desktop muon counter project combines all of these aspects into a unified program and delivers a useful physics tool.

Similar devices are used in particle physics experiments to identify well-reconstructed muon track samples for detector calibration. An existing example includes the ``muon cubes'' that were installed in the MiniBooNE neutrino experiment at Fermilab \cite{miniboone}. These optically isolated scintillator cubes were used in combination with a set of scintillation counters located above the detector to accurately track and tag stopping muons. Another similar project, ArduSiPM\cite{Valerio}, was developed as a cost-effective way to read out information from silicon photomultipliers (SiPMs) and has found applications in radio-guided surgery as a $\beta$-probe~\cite{nature}.

The original desktop muon counter, produced at MIT, was a prototype for a subdetector of PINGU, an upgrade to the IceCube Neutrino Observatory, located at the South Pole \cite{IceCube}. In PINGU, optically isolated cubes located throughout the detector could provide well-defined hits on a set of muon tracks, allowing tests of track reconstruction.  Thus, development of this detector is a realistic exercise for students who intend to participate in particle physics and astrophysics experiments in the future.

\section{The components of this project}
The device consists of a small slab of solid scintillator instrumented with a silicon photomultiplier (SiPM) to detect scintillation light. It is contained within a light-tight aluminum enclosure machined by the student. This connects to a readout box consisting of electronics that register the time of the event, count number, peak amplitude, and dead time. The threshold for a signal to trigger the data acquisition can be tuned in the microcontroller (Arduino) software. We discuss the individual components for the project below as well as provide pictures of each component in the supplementary material (pictures/).

\subsection{Scintillator \label{scint}}

Scintillators emit light when a charged particle passes through them and deposits a fraction of its initial energy due to electromagnetic interactions. The amount of emitted light is related to the energy of the incident particle and the distance the particle traverses through the scintillator. In this case, the scintillator responds to this energy because the plastic is doped with a fluorescing agent that glows very slightly when some kinetic energy is transferred to the fluorescing molecules. Within nanoseconds, the de-excitation of these fluorescing molecules produces visible light, typically in the 300 to 600~nm wavelength range, that travels through and exits the scintillator.

Scintillators come in a number of forms other than solid plastic.    For example, there are inorganic solid scintillators and liquid scintillators. However, we recommend a plastic scintillator for this project because it is inexpensive and easy to handle.

One can purchase new scintillators or use old scintillator paddles, as long as they are sufficiently thick (in the described design, we use a 5$\times$5$\times$1~cm$^3$ solid piece of scintillator).   Because the detector is very small, it is not a problem if the used scintillator has some minor damage. If your department has no used scintillators available, we have found used paddles for sale for a very reasonable price on eBay.  New scintillators can be purchased if necessary from companies such as Saint Gobain \cite{stgobain} or Eljen \cite{eljen}.

The scintillator slabs may have to be machined to size on a mill and then polished in order to make the faces optically transparent. Polishing the scintillator improves the photon collection efficiency of the SiPM by increasing the optical transparency at the interface between the SiPM and the plastic scintillator. It also promotes total internal reflection off the walls of the scintillator, thus increasing the overall number of photons. We also wrap the plastic scintillator in reflective foil to increase the number of photons reflected back towards the SiPM face. We use optical gel to match the refractive indices of the plastic scintillator and the protective layer on the SiPM's photocathode to increase efficiency. 

\subsection{Silicon Photomultipliers}

The light emitted when a particle travels through the scintillator must be observed using a light collection device. Traditionally, one attaches the scintillator to photomultiplier tubes (PMTs) like those found in the portable muon detector project $\mu-Witness$ \cite{witness}. These are large, require high voltages, and are expensive.  In this case, we use a single SiPM, since it requires only a low reverse bias voltage (positive voltage to the cathode, negative voltage to the anode), has a peak sensitivity in the blue region where the majority of scintillators emit most of their light, and is only a few millimeters thick with a cross-sectional area equal to the size of the photocathode. The low reverse bias means that we can use an inexpensive DC-DC boost converter to power the circuit. 

A SiPM consists of a large number of microcells, each composed of silicon P-N junctions. Electrons migrate into the P-side and holes migrate into the N-side. This creates a region known as the ``depletion region," where the electrons and holes eliminate through recombination. When a photon traverses the depleted region, it can deposit sufficient energy to an electron in the valence band to move it to the conduction band, thereby creating a current. Biasing the P-N junction increases the depletion region and creates an electric field $>$5$\times$10$^5$~V/cm. When a charge carrier accelerates through this field, it can gain sufficient kinetic energy to ionize the surrounding atoms through impact ionization. This creates an avalanche of electron-hole pairs, which can have a gain as high as 1$\times$10$^7$. Thus, a single photon producing a single electron-hole pair can generate a very large, measurable signal.

SiPMs have a very high dark-noise rate. These are signals that occur randomly when thermodynamic processes in the silicon generate an electron-hole pair that proceeds to avalanche. This signal is indistinguishable from that produced by a single photon. Therefore, it is essential that a muon passing through the scintillator produces enough visible photons within a short time period so that the resulting signal is much larger than the background noise. Our counter is designed with this goal in mind.

The detector described in this paper uses a 6$\times$6~mm$^2$ C-Series 60035-SMT (surface-mount technology) SensL SiPM~\cite{sensl}. These SiPMs have a breakdown voltage of roughly 24.7~V and can sustain an overvoltage of up to 5.0~V. The cost of SiPMs drops rapidly with the number that are purchased. Thus, it is most cost-effective for a department to buy a relatively large number of SiPMs for multiple classes at once or to purchase them in conjunction with other classes or experiments. At the time of writing this paper, the unit price of a bulk purchase of 100 SiPMs was below \$50/SiPM. This cost represents approximately half of the total cost of construction of the desktop muon counter.

\subsection{Electronics components}

There are two PCBs, each with surface mount components that the students will install. 
The simplest of the two boards is used to mount the SiPM and provide bias filtering. This PCB is mounted directly onto the plastic scintillator by means of two No. 0 1/4$''$ screws to maintain pressure on the SiPM face thus ensuring good optical contact between the photocathode area and the plastic scintillator. The second PCB contains the main electronics used to amplify and shape the signal from the SiPM such that it can be measured by the microcontroller. It also filters and regulates the voltages used in the detector. The amplification and shaping of the waveform is accomplished using dual rail-to-rail input and output operational amplifiers (op amps), whose functions are described in detail in Sec. \ref{sec:electronics}. We use an inexpensive 16~MHz Arduino Nano ATmega328 as a microcontroller and read the data out to a 0.96-inch OLED screen through a mini-USB cable to a computer. The code necessary to run the Arduino (Arduino/Arduino\_code) as well as a list of the required libraries (which all can be installed in the Arduino integrated development environment, IDE) are supplied in the supplementary material  (Arduino/Arduino\_library\_list). We also provide a Python script to run on a computer to log the data (Arduino/Import\_data.py). The Python script requires that the students install the pyserial module. Students are asked to design their own program to analyze the data.

\subsection{The light-tight enclosure} \label{sec:box}

The plastic scintillator and SiPM circuit are mounted in a light-tight aluminum enclosure. The enclosure not only keeps light from the scintillator inside, but it also protects against photons entering from outside.  This prevents environmental noise from producing false signals. However, the metal box has to be penetrated to power the SiPM and to send the signal from the SiPM to the electronics box. Commercial DC jacks and BNC (Bayonet Neill-Concelman) connectors are not quite light-tight; therefore, as a precaution, it is recommended that the plastic scintillator and SiPM be wrapped in black electrical tape. This component can also be completed externally at a machine shop, simply by providing the supplied drawings in the supplementary material.

\subsection{Cables and electronics case}
The signal from the SiPM is transmitted out of the light-tight enclosure to an electronics case via a 6$''$ BNC cable. Here, the students are asked to manufacture their own BNC cable and check the continuity of their connections. Since BNC cables are so prevalent in physics labs due to their robust coaxial design and RF shielding characteristics, understanding how to make your own or repair a noisy cable is an important skill. We use a simple 2.1$\times$5.5~mm$^2$ DC cable to power the SiPM circuit. 

The electronics case houses the main PCB. There are many design options for the electronics case. It is crucial to make sure that the electronics can be securely mounted with enough room to allow for the final soldering of connections. There should be at least an internal volume of 3$\times$4$\times$1 in$^3$ to accommodate the electronics. In the designs shown in Fig.~\ref{fig:array}, we use repurposed Ethernet switching cases. 

\section{Manufacturing the components}
This section describes the machining, 3D-printing, and PCB board manufacturing required for each component. Further technical material (CAD drawings, files, programs, and documentation) on the manufacturing can be found in the supplementary material and will be referenced as needed.

\subsection{Machining the light-tight enclosure}
\begin{figure}[h!]
 \centering
\includegraphics[width=0.8\columnwidth]{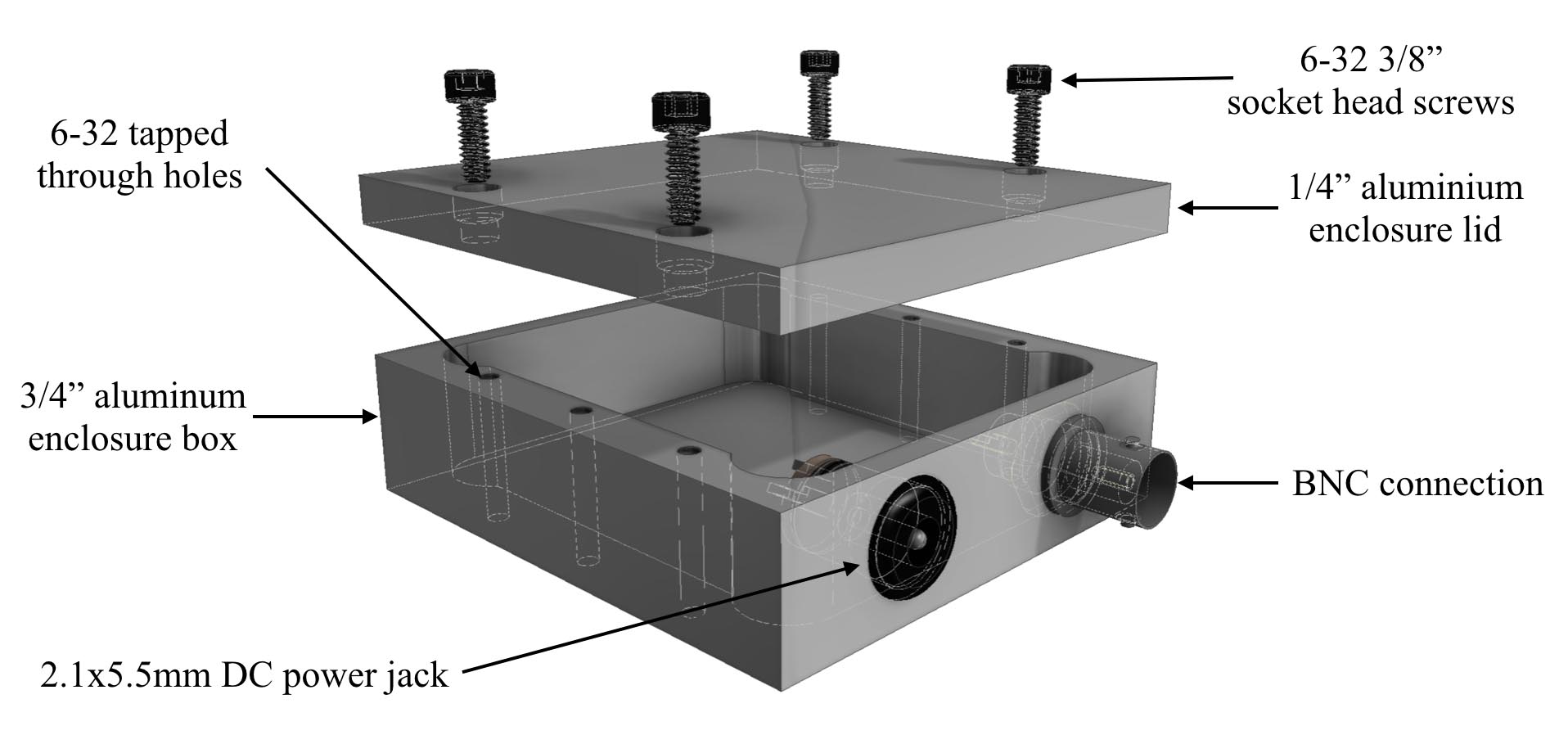}
  \caption{A rendering of the light-tight enclosure. Computer-aided design files are found in the supplementary material~\cite{sup}.}
  \label{fig:metal_box}
\end{figure}

The light-tight enclosure, shown in Fig.~\ref{fig:metal_box}, houses the plastic scintillator, SiPM, and SiPM PCB. The enclosure consists of an aluminum box and lid, which are mated together with four 6-32$\times$3/8 socket head screws.

The enclosure box is made from a stock piece of 2.75$\times$3.0$\times$3/4 in$^3$ 6061 aluminum. A pocket is carved to a depth of 0.65 inches using a 2-flute, 1/2 inch end mill into the aluminum on a programmable mill. The pocket can also be carved manually but it will take more time. A .dxf file containing the outline of the pocket can be found in the supplementary material~(machining/box\_CNC.dxf). Six 6-32 through holes are tapped on the outer edges of enclosure. The outer four holes are used to secure the enclosure lid to the box, while the middle two are used to secure the enclosure to the electronics case. We use through holes here, instead of blind holes, to allow the chips to fall through when tapping. This decreases the chances of breaking the tap. After the holes are tapped, the top surface of the box and bottom surface of the lid should be faced on the mill in order to provide a light-tight seal. We use a 3-inch fly cutter to face the surfaces in a single pass.

There are two holes on the back end of the enclosure box. The smaller of the two measures 3/8 inches in diameter and is used for the female BNC nut bulkhead connector. The larger, a 2.1$\times$5.5~mm$^2$ DC power jack, is used to supply the 29.4~V required to power the SiPM circuit.

The enclosure lid is made from a stock piece of 2.75$\times$3.0$\times$1/4 in$^3$ 6061 aluminum plate. The bottom side must also be faced to ensure the box is light-tight when closed. Four countersunk through holes for the four 6-32 socket head screws are drilled on the outer edge to line up with the four holes in the enclosure box. The CAD drawings (machining/box.pdf,lid.pdf)
and files for programming the mill (machining/box\_CNC.dxf) are provided in the supplementary material. We recommend that the students use the mill for the entire manufacturing process of this component. This will help ensure that the holes are properly aligned and the edges are square. Further, after the enclosure is assembled, all six faces can be faced to provide a smooth, polished final finish. 

\subsection{Polishing the plastic scintillator}
The plastic scintillator can be cut to the approximate size using a band saw and then side-milled using a 2-flute, 1/2 inch endmill at approximately 1200 rpm to the final dimensions. Our final size measured 50$\times$50$\times$10 mm$^3$. The faces that were already transparent did not need to be machined. We found that a very shallow final pass on the mill provided a very smooth, although visibly murky finish. Two holes are then drilled into the face of the plastic scintillator to mount the SiPM PCB. These holes must be drilled at low speeds (less than 100~rpm). High-speed drilling creates too much heat and the scintillator will often crack as it cools. The holes are spaced 40~mm apart, to a depth of 1/4 inches, using a number 54 drill bit. These holes are sufficiently large that a No. 0 machine screw will be able to self-tap into the plastic scintillator. 

Upon finishing the machining of the plastic scintillator, the surfaces must be polished to provide an optically transparent face for the SiPM. We have experimented with two different methods to polish the plastic scintillator. The simplest but most time-consuming method to improve the optical transmission through the machined face is to use incrementally finer grit sand paper and then polish the final surface on a polishing wheel. The second method is to ``flame" polish the machined surface using a hot air gun from a soldering station. This is a quick maneuver in which we heat the surface of the plastic scintillator just long enough for it to become clear (roughly several seconds). As the solid scintillator melts and resolidifies, the surface becomes optically transparent. Flame polishing introduces new stresses in the plastic scintillator, and therefore all machining, including drilling the holes to mount the SiPM PCB, must be completed prior to this step. Flame polishing has the added benefit that the machine surface does not need to be perfectly flat. We found that flame polishing the surface even after cutting the scintillator on the band saw and filing the surface flat provided adequate results. This may be preferable for students attempting to build a device without access to a mill.

 \begin{figure}[h!]
 \centering
\includegraphics[width=0.90\columnwidth]{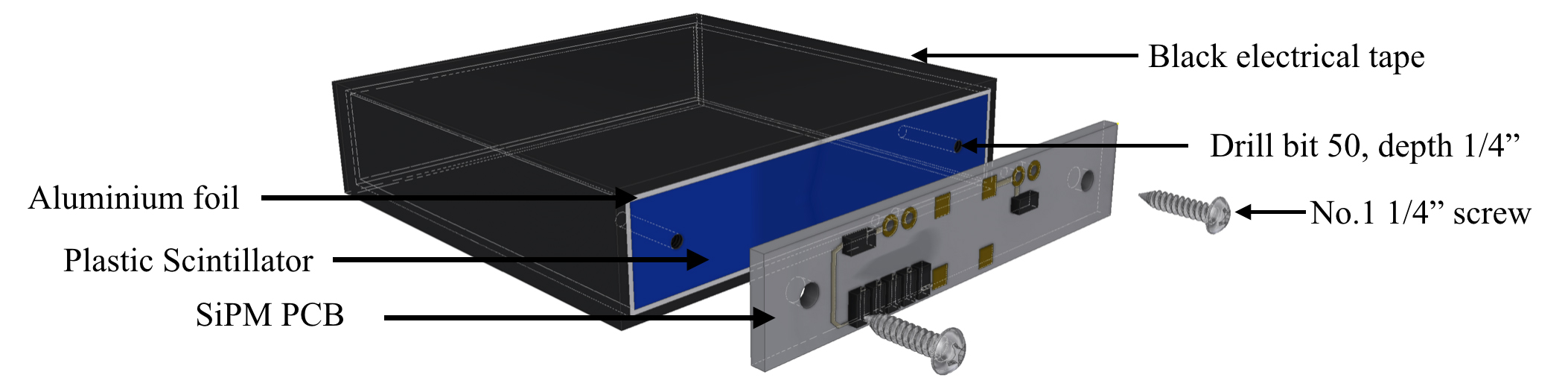}
  \caption{A rendering of the plastic scintillator and SiPM assembly. The plastic scintillator is wrapped in aluminum foil and secured in place with black electrical tape.} 
  \label{fig:foil}
\end{figure}

\subsection{Machining the electronics box}
Here we provide an example of the electronics box design for the muon detector using a repurposed Ethernet switch box. Ethernet switch boxes tend to be correctly sized, have very few internal components, and are therefore a relatively inexpensive enclosure. Our chosen Ethernet switch box also came with the required 5~V, 1.0~amp wall adapter supply. The box must have adequate volume to support the main PCB board, which is roughly 3$\times$4$\times$1~in$^3$. 

Several through holes in the box need to be machined to mount the other components. When drilling through plastic, it is useful to use a sharp, 60$^{\circ}$ point angle drill bit to eliminate edge chipping and chip wrapping. In our case, the top of the electronics box is used to mount the light-tight enclosure and the OLED screen case. Both are secured in place using 6-32 machine screws. A separate hole, directly underneath the OLED case, is used to run the OLED cables. A rectangular hole for the mini-USB connection is easily machined using a rotary drill but can also be done using a mill. It was also necessary to attach an aluminum plate on the back of the electronics box to mount a reset button, BNC connector, and DC barrel jack to connect to the light-tight enclosure. Photos of the electronics box can be found in the supplementary material.

\subsection{OLED screen case}
The 0.96-inch OLED screen readout is mounted on a 3D-printed frame to secure it to the electronics box to protect it from damage. The file required to print the case is found in the supplementary material (machining/OLED\_case.stp). It is printed face down and requires approximately 0.5 cubic inches of printing filament, including scaffolding. The screen is held in place by gluing the front end of the OLED PCB board to the interior of the case. 

 \begin{figure}[h!]
 \centering
\includegraphics[width=0.80\columnwidth]{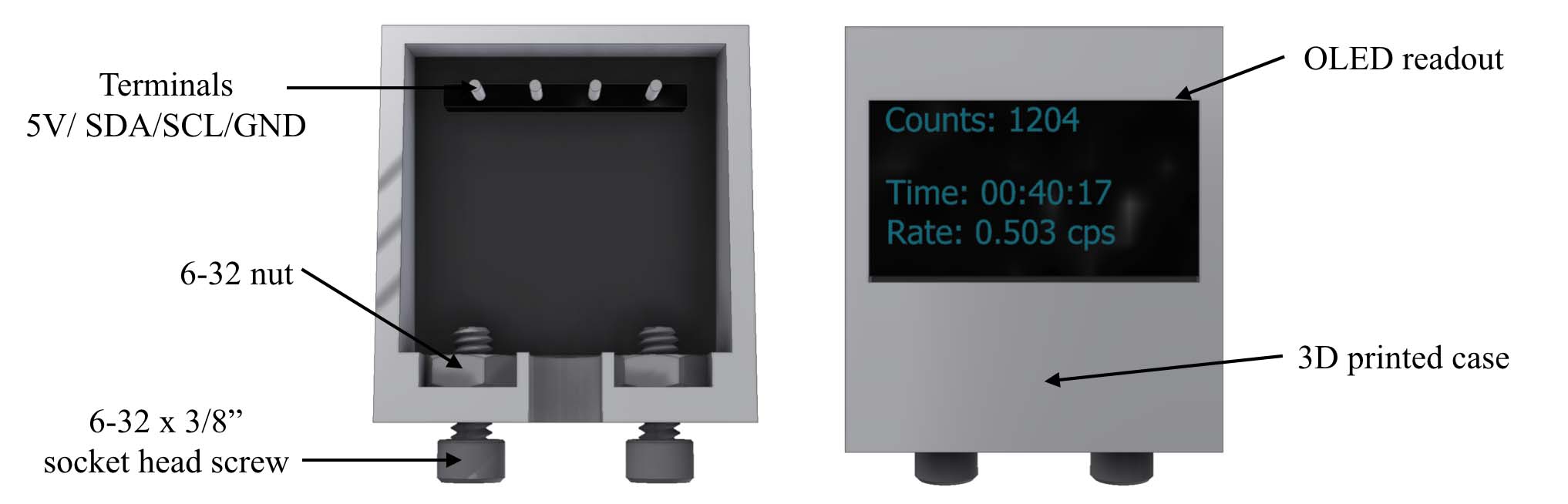}
  \caption{Front and back view of the 3D printed OLED screen case.}
  \label{fig:elecs}
\end{figure}
 
\subsection{Circuit boards}
The two PCBs must first be manufactured from an electronics company. For example, Elecrow.com~\cite{elecrow} provides manufacturing of custom PCBs at a reasonable cost. The necessary files (PCB/MAIN\_PCB.zip and PCB/SiPM\_PCB.zip) are provided in the supplementary material. The color of the SiPM PCB should be white to improve reflection in the scintillator. The SiPM PCB measures 50$\times$10$\times$1.6~mm$^3$, while the main PCB is 50$\times$50$\times$1.6~mm$^3$. Both circuits were designed in KiCAD~\cite{KiCAD} as two-layer boards. A rendering of the PCBs can be found in the supplementary material.

The required electronic components for populating the PCBs can all be purchased from either Amazon~\cite{Amazon} or Digi-Key~\cite{DigiKey}. The reference locations for all the surface mounted components are listed in PCB/SMT\_reference.xlxs.

Both PCBs were designed using 0805 surface mount components (0.08$\times$0.05~in$^2$ or 2.0$\times$1.2~mm$^2$). These are small but relatively easy to manipulate with a good pair of tweezers. We suggest searching online for a video to learn the proper techniques for using surface mount components.  Adafruit provides an excellent tutorial here in \cite{tutorial}, but there are hundreds to choose from. The PCBs should be populated using a fine tip soldering iron but reflow solder and an oven can also be used. If one uses an oven, consult the SiPM documentation for the temperature profile.

The SiPM PCB has a silkscreen outline where the SiPM is to be mounted. We found the best way to solder the SiPM in place is to first put a bit of solder on one of the PCB pads, then line the SiPM up exactly with the silkscreen and solder a single pin in place. Once the SiPM is properly aligned between all the pads, the other pins can be soldered in place. Although the SiPM PCB has four pads for the SiPM footprint, it is only necessary to solder pads 1 and 3. Pin 1 on the SiPM can be identified by looking carefully at the number of metal legs on each corner. Pin 1 has three legs, all the others have two.

The main PCB board, can be assembled according to the component reference list in PCB/SMT\_reference.xlxs. The outline on the silkscreen provides the location of where the larger components are to be mounted. The IN+ and IN- terminals labeled on the DC-DC boost converter silkscreen are used to solder the booster onto the corresponding main PCB pads. Leads from the OUT+ and OUT- on the DC booster must be connected to the DC power jack on the electronics case. The potentiometer on the DC booster can be used to adjust the output voltage. It should be set such that the output voltage is +29.4~V.

There are several header pins used to attach the reset switch, OLED screen, and boost converter and to receive the signal input from the SiPM. These should be all be attached during the assembly of the detector, once the main PCB has been mounted to the electronics box. 

\section{Assembling the detector}

 \begin{figure}[h!]
 \centering
\includegraphics[width=1\columnwidth]{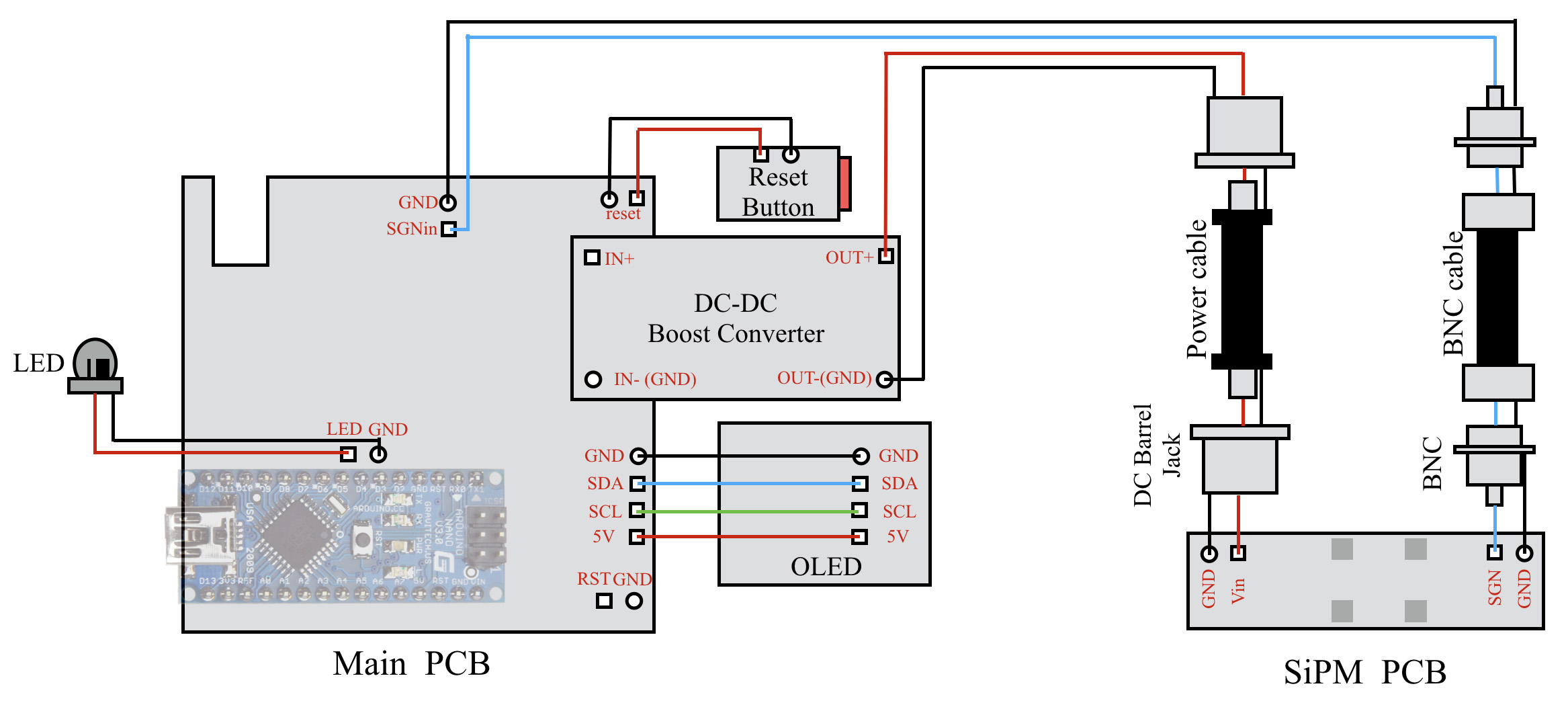}
  \caption{A wiring diagram of the detector components.}
  \label{fig:wiring}
\end{figure}

The SiPM PCB is fixed to the face of the plastic scintillator using the two No. 0 1/4 inch screws. A small amount of optical gel is used to interface the SiPM to the plastic scintillator. The screws provide enough pressure to remove air bubbles from the optical gel, but not enough pressure to bend the PCB or damage the SiPM. A reflective foil is then wrapped around the plastic scintillator and held in place using electrical tape. It is crucial that the reflective foil not come into contact with any part of circuit. Taping around the SiPM PCB and the plastic scintillator will also improve the overall light-tightness.

\begin{figure}[h!]
 \centering
\includegraphics[width=1\columnwidth]{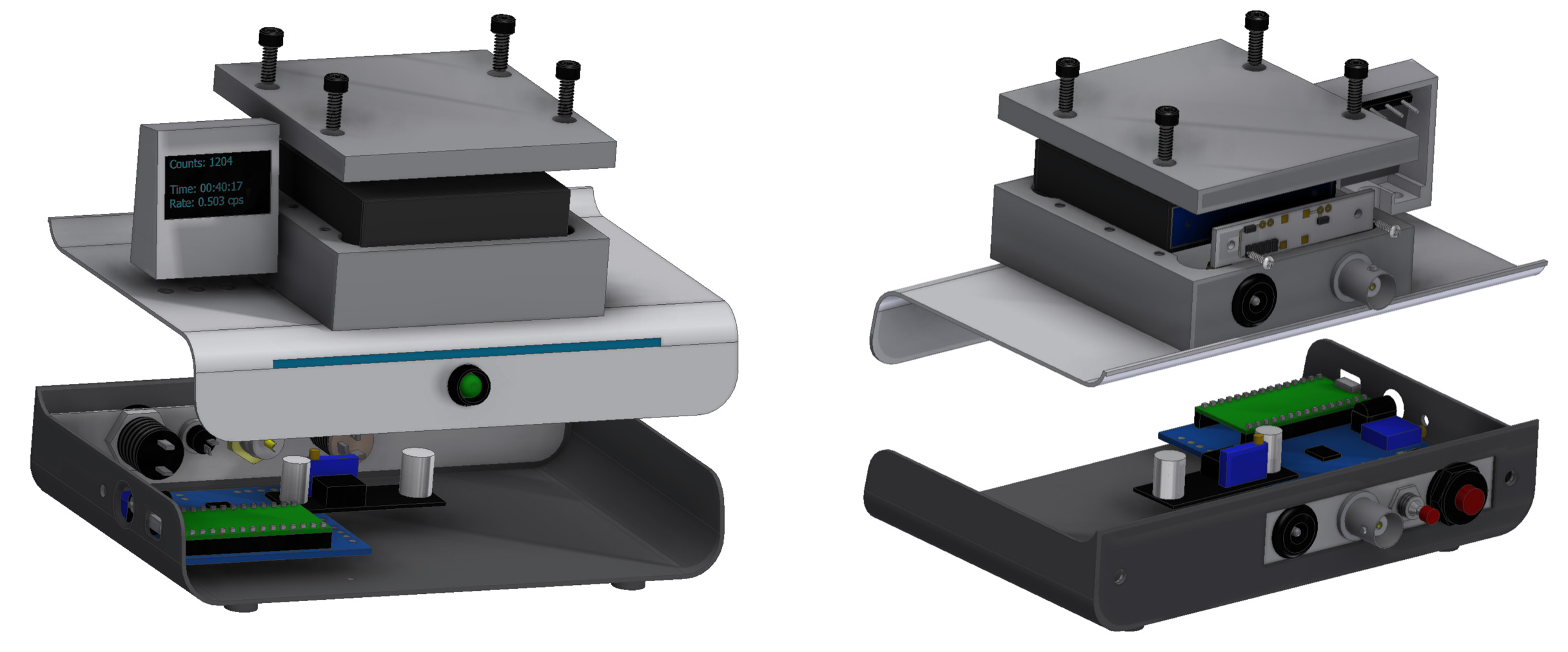}
  \caption{The complete assembly of the desktop muon detector.}
  \label{fig:assembly}
\end{figure}

The scintillator is then inserted into the light-tight enclosure box. Leads from the signal connection (labeled SNG) and ground connection (GND) are connected to the female BNC connector, and the V$_{\mathrm{IN}}$ and GND are connected to the 2.1$\times$5.5 mm$^2$ DC power jack according to the wiring in diagram Fig.~\ref{fig:wiring}. The aluminum enclosure lid can then be secured using the four 6-32$\times$3/8 inch screws and mounted onto the electronics box (see Fig.~\ref{fig:assembly}).

Once the assembly of the light-tight enclosure has been completed, it can be tested using an oscilloscope and variable 30~V power supply. Supplying the DC jack connection with 24.7 to 29.5~V should create positive pulses with an amplitude of 10-100~mV that will exponentially decay in roughly 0.5~$\mu$s when a muon passes through the scintillator. At sea level, one should expect to see roughly one pulse per cm$^2$ per minute due to cosmic ray muons. Low-amplitude pulses may indicate that the enclosure is not light-tight, the face of the SiPM is not in good contact with the plastic scintillator, the scintillator face was not adequately polished, or the SiPM face could be damaged. A larger count rate than roughly one pulse per cm$^2$ per minute can be explained by the background contamination of radioisotopes in and around the enclosure. We found that gamma rays from the outside environment will also penetrate the aluminum enclosure, but $\beta$ and $\alpha$ are significantly attenuated.

The main PCB is secured to the electronics case through the two holes on the bottom of the case. For our design, we required a 1/4$''$ standoff under the main PCB to bring it to the right level for the DC power jack port. Once the main PCB has been secured to the electronics case, the leads for the reset switch, OLED screen, DC-DC boost converter, power for the SiPM, and the signal input from the SiPM can all be attached according to Fig.~\ref{fig:wiring}.

The OLED screen is fixed to the 3D-printed screen case with a small amount of epoxy, and leads are to be connected to the four terminals on the back. These leads are then fed through a hole on the top of the electronics case to the main PCB and are wired according to Fig.~\ref{fig:wiring}. 

The final step in the assembly process is to upload the Arduino code using a mini-USB cable. This requires the student to install the Arduino IDE and the libraries listed in the supplementary material (Arduino/library\_list.pdf). The libraries are used to communicate with the OLED screen and Arduino timer interrupts. All the libraries can be installed through the Arduino IDE except for OzOLED, which is referenced in the supplementary material. The particular Arduino Nano that we purchased required a specific driver in order to communicate with the Mac OS. The manufacturer should provide a link to the location of their driver files.

Powering the device requires either the USB or a 5~V power cable of at least 250~mA to be connected to the main PCB. The full detector consumes less than 1~W of power. If running the detector off a battery or at high count rates, it is recommended that the OLED and LED are turned off. This can be done by changing the booleans ``OLED" and ``LED" to 0 in the Arduino code.

\section{The electronics circuitry}\label{sec:electronics}
 
 \begin{figure}[h!]
 \centering
\includegraphics[width=1.0\columnwidth]{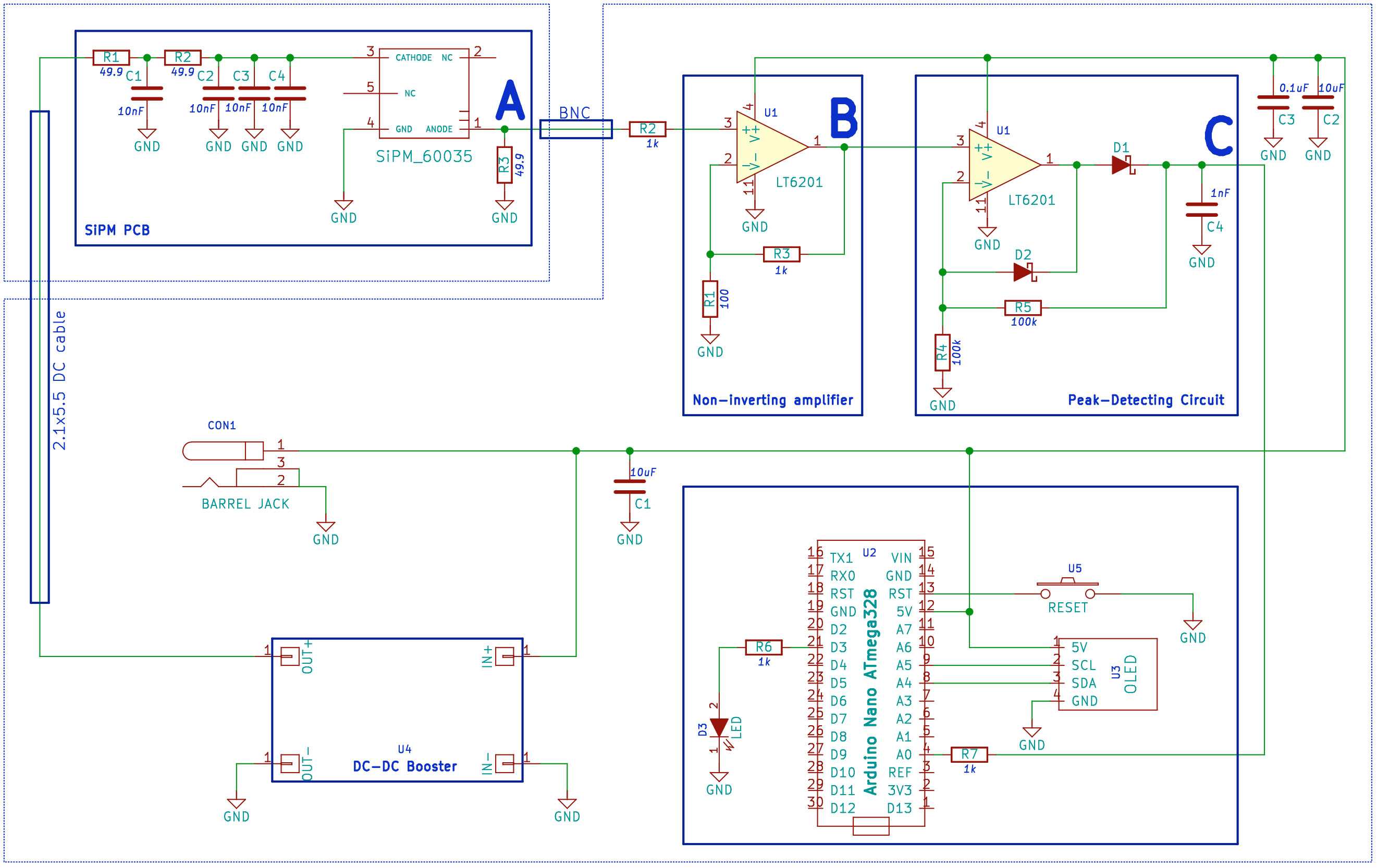}
  \caption{The complete circuit diagram. The position ``A" represents the signal from the SiPM; ``B," the signal from the amplifier; and ``C," the signal from the peak detector.}
  \label{fig:electronics}
\end{figure}

The complete electronics circuit is shown in Fig.~\ref{fig:electronics}. The SiPM circuit, outlined in blue on the left, is mounted on the plastic scintillator in the light-tight enclosure, while the rest of the circuitry is contained on the main PCB. In the following, we will describe the general principle behind how the circuit was designed and then give a more in-depth explanation of the various components and their respective functions.

\subsection{The electronics overview}
\begin{figure}[h!]
 \centering
\includegraphics[width=0.7\columnwidth]{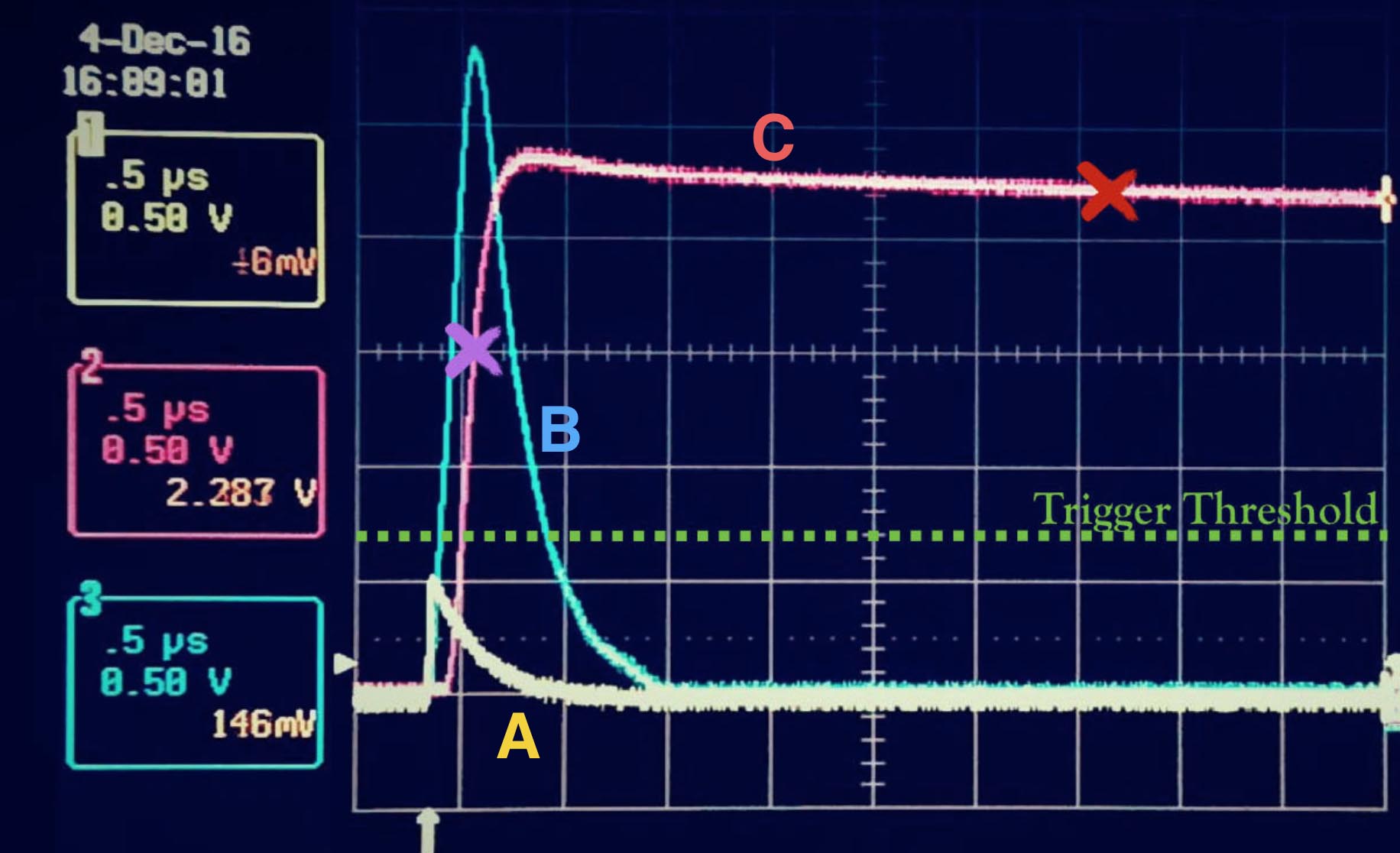}
  \caption{Traces on an oscilloscope at the various parts of the circuit.  The software trigger threshold, shown as a dotted green line, is set in the Arduino code. The waveform labeling (A, B, C) corresponds to the parts of the circuit in Fig.~\ref{fig:electronics} that we are measuring. The red ``X'' represents the first Arduino sample.}
  \label{fig:principle}
\end{figure}

The general principle behind how we are measuring the signal from a muon interaction is shown in Fig.~\ref{fig:principle}. In this figure, there are three waveforms, labeled ``A," ``B," and ``C," that correspond to the positions labeled similarly in Fig.~\ref{fig:electronics}. A muon-induced photo-avalanche in the SiPM will create a positive pulse, whose width is $\mathcal{O}(0.5~\mu$s) and height is typically between 10--100~mV. This pulse is sent through a noninverting amplifying circuit. Here, the pulse is amplified by approximately a factor of six and passed to a peak-detector circuit that outputs a pulse that rises to the peak of the amplified pulse but decays slowly over a period of roughly 100~$\mu$s. The Arduino samples the decaying pulse and uses this information to calculate the initial pulse amplitude.

\subsection{The SiPM circuit}
The DC-DC boost converter circuit, shown in the bottom left of Fig. \ref{fig:electronics}, is an off-the-shelf device that converters the 5~V input to the SiPM breakdown voltage (approximately 24.7~V) plus an overvoltage. We have chosen to operate the SiPM at an overvoltage of 4.7~V, which improves the low-level signal response but in return also increases the dark rate. The potentiometer on the DC-DC boost converter is used to bring the potential between the OUT+ and OUT- pins to +29.4~V.

There is a 10~uF electrolytic capacitor (C1 in Fig.~\ref{fig:electronics}) at the output of the power supply, which is known as a bypass capacitor (decoupling capacitor). This capacitor has two main purposes: to locally store energy for when the SiPM discharges and to act as a filter to decouple noise generated by the power supply.  The internal impedance of the capacitor causes it to act as a low-pass filter, letting low frequencies through and suppressing high frequencies. 

Prior to the SiPM, the bias voltage is sent through a series of low-pass filters. A low-pass filter attenuates frequencies higher than the ``cut-off" frequency. The cut-off frequency is defined as the frequency at which the signal intensity drops to 63.2\% its original value. Schematically, it is represented by a resistor followed by a capacitor to ground, and the equation representing the cut-off frequency is $\mathrm{f_{cut}} = \frac{1}{2 \pi R_1 C_1}$. Two low-pass filters can be seen on the left of the SiPM circuit.

We positively bias the cathode (Pin 3) of SiPM to +29.4~V, while Pin 4 is connected to ground (optional) and Pins 2 and 5 are left open. The resistor to ground on the anode (Pin 1) of the SiPM is called a ``pull-down resistor," which holds the line at ground when there is no signal. An induced pulse in the SiPM circuit is then sent to the amplifying circuit.

\subsection{The amplification circuit}
\begin{figure*}
    \centering
    \begin{minipage}{.45\linewidth}
        \centering
        \includegraphics[width=1\columnwidth]{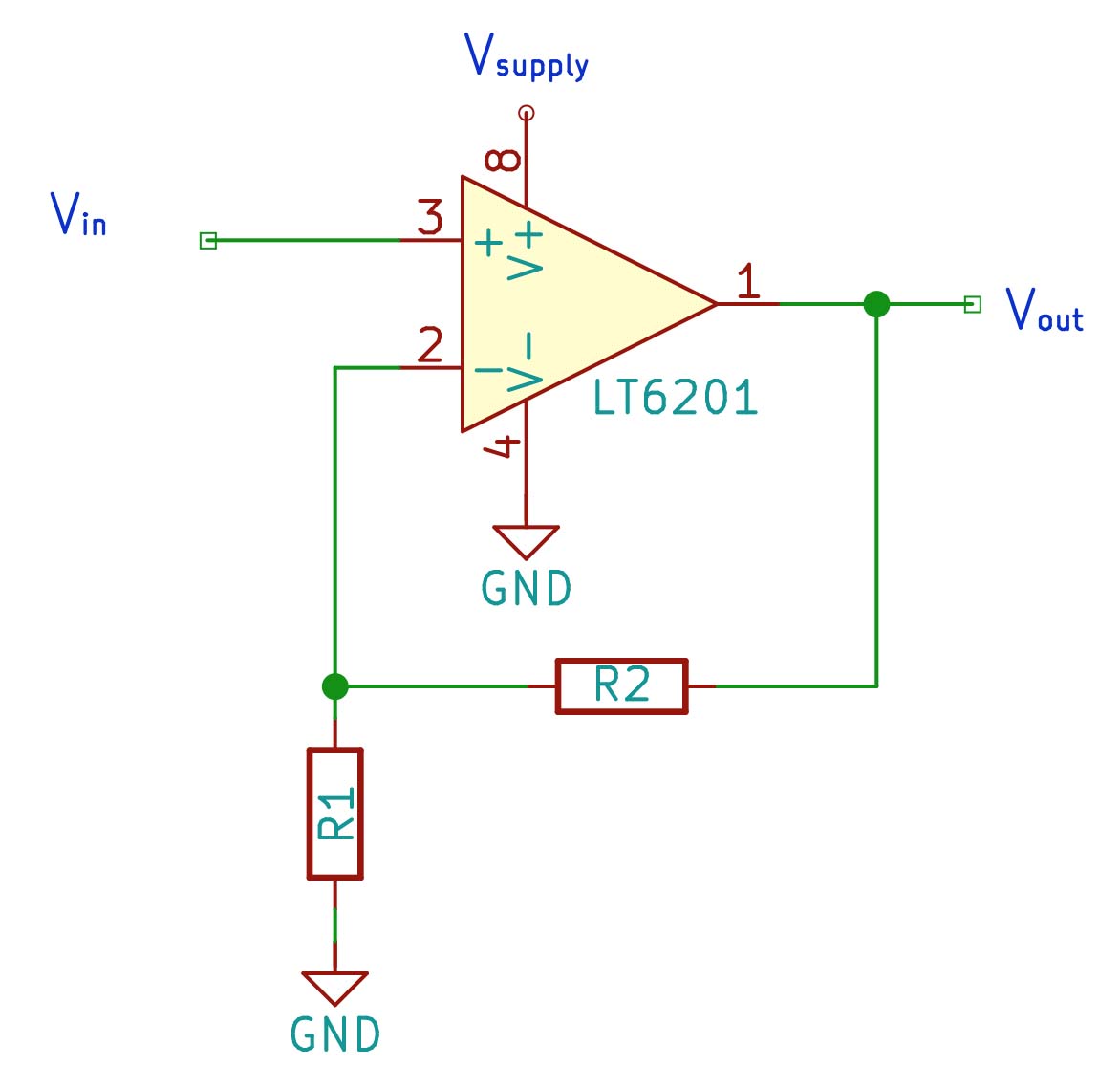}\label{fig:amp}
    \end{minipage}%
    \hfill%
    \begin{minipage}{.45\textwidth}
    \centering
        \centering
        \begin{equation}\label{equ:amp}
            \mathrm{V_{OUT}} = \mathrm{V_{IN}}(1+\frac{R_2}{R_1})
        \end{equation}
    \end{minipage}\\[-7pt]
    \begin{minipage}[t]{1\linewidth}
        \caption{Single supply noninverting op amp circuit and associated gain equation.}
    \end{minipage}%
    \hfill%
\end{figure*}

This part of the circuit is know as a ``single-supply noninverting operational amplifier circuit." It takes the positive pulse from the SiPM, V$_{\mathrm{IN}}$, and amplifies it to a positive pulse, V$_{\mathrm{OUT}}$, according to Eq. \ref{equ:amp}. The ``single-supply" refers to the fact that we are supplying +5~V to the positive rail, V$_{\mathrm{+}}$, and setting the negative rail, V$_{\mathrm{-}}$, to ground. An in-depth description of the operational principles behind op amps can be found in Ref. \cite{horrowitz}.

Using the resistor values in Fig. \ref{fig:electronics} (R1 = 100~$\Omega$, R2 = 1~k$\Omega$), the ratio between output voltage and input voltage in Eq. \ref{equ:amp} indicates we should expect an amplification, or gain, of 11. However, due to the limited frequency response of the op amp, this is not quite the case. The circuit was designed using an op amp (LT6201) with a gain-bandwidth product of 145~MHz, which, at a gain of 11, gives a bandwidth of approximately 13.2 MHz. Since the rise time of a signal from the SiPM is a few tens of nanoseconds, we expect this high-frequency component to be attenuated. The measured peak-to-peak amplification was found to be approximately 6, as indicated by the traces in Fig.~\ref{fig:principle}.

\subsection{The peak detector circuit}
The purpose of the peak detector circuit is to detect the amplitude of the amplified pulse and hold the voltage at that level for a sufficient time such that the Arduino can measure it, then decay and wait for the next pulse. 

Fig. 10 shows the electronic schematic for our peak detector circuit. This circuit was modified from a circuit found in \cite{peak}. Once a pulse from the amplifying circuit enters the noninverting input of the op amp (+), the Schottky diode D2 becomes forward-biased and allows the op amp to charge the sampling capacitor C1. While charging, there is an unavoidable leakage current through the resistors R1 and R2 to ground. However, these resistors were chosen to be large enough so that this is negligible. When the pulse from the amplifying circuit subsides, D2 becomes back-biased and forces C1 to discharge through R2. The current will then flow to ground via two different paths depending on the voltage on C1.

If there is a large voltage on C1 (greater than the forward voltage drop on D1), D1 becomes forward-biased and will allow current to flow to the output of the op amp, which is now sitting at the negative rail, in our case it is ground. The decay time associated with this is then R2$\times$C1. If the voltage on C1 is smaller than the forward voltage drop on D1, the diode will be back-biased and current will flow through the series of resistors R1 and R2. The decay constant associated with this is (R1+R2)$\times$C1. This bifurcation was found to greatly improve the response of the circuit to very small and very large incoming pulses. The decay time was found to be sufficiently long for the Arduino to sample the pulse multiple times.

Since the output of the op amp can only be driven to 4.78~V (with 5~V supplied to the positive rail) and the voltage drop across D2 is approximately 0.4~V, the maximum output voltage now becomes approximately 4.28~V. We have specifically chosen the diodes to minimize the forward voltage drop, thus allowing us to measure a higher possible voltage. 

\begin{figure*}
    \centering
    \begin{minipage}{.45\linewidth}
        \centering
        \includegraphics[width=1\columnwidth]{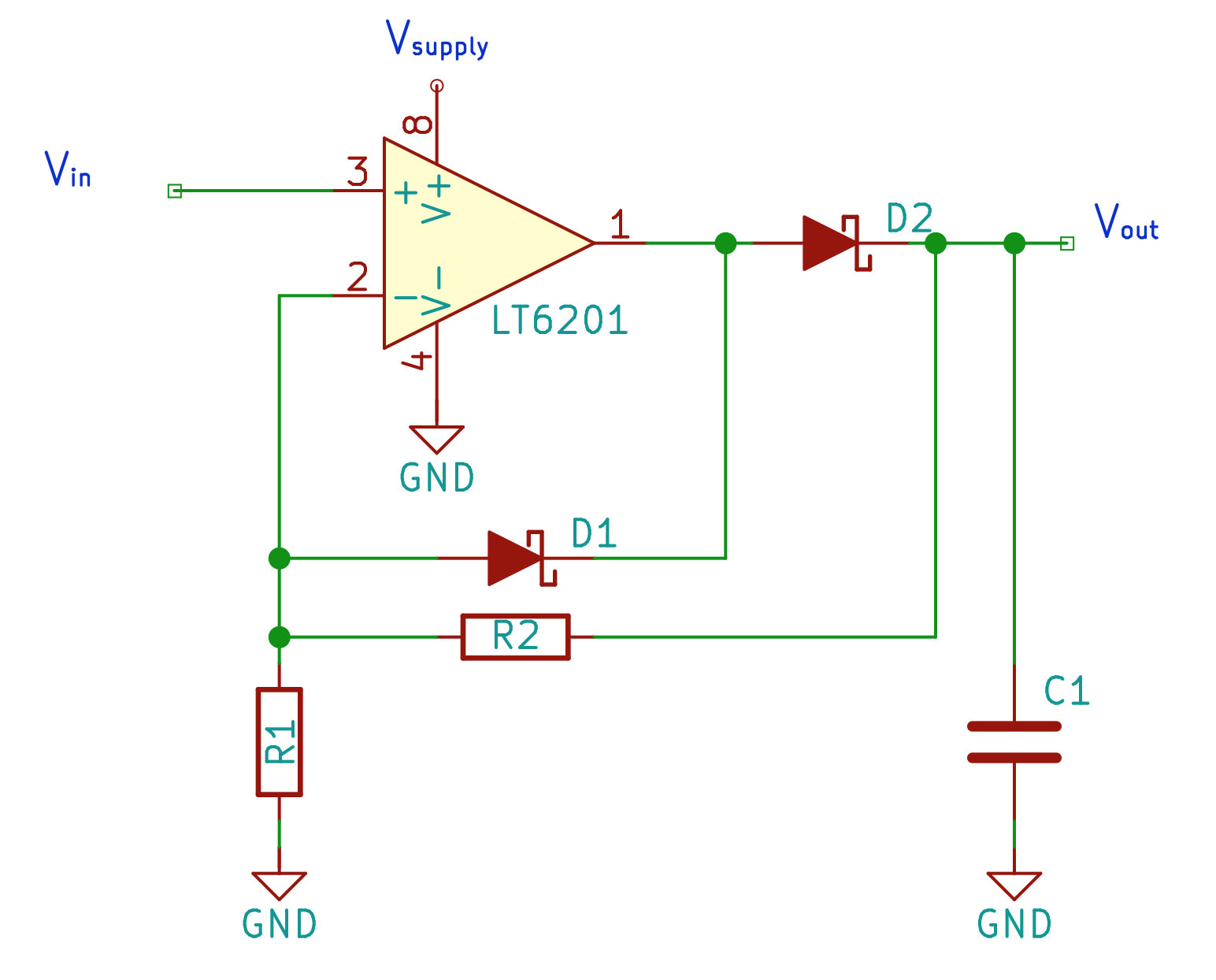}\label{fig:pd}
    \end{minipage}%
    \hfill%
    \begin{minipage}{.45\textwidth}
    \centering
        \centering
        \begin{equation}\label{eqn:time}
            \mathrm{\tau} = R \times C
        \end{equation}
    \end{minipage}\\[-7pt]
    \begin{minipage}[t]{1\linewidth}
        \caption{The peak detector circuit. We have selected R1~=~R2~=~100k$\Omega$ and C1 = 1~nF.}
    \end{minipage}%
    \hfill%
\end{figure*}

\subsection{The Arduino circuit}
The analog Arduino pin, A0, monitors the output waveform from the peak detector. If the voltage rises above the trigger threshold, the Arduino makes several measurements to calculate the original pulse height. The analog pins have a 10-bit resolution that ranges from 0--5~V. This corresponds to a voltage measurement resolution of approximately 5~mV. Although the Arduino has a clock speed of 16~MHz, we cannot sample the waveform at this rate. We found that, using a prescaler of 4, the sample frequency was measured to be approximately 172~KHz (5.8~$\mu$s per sample). This is sufficient for our purposes but limits the triggering signal resolution to a few microseconds. 

Since the trigger sample (red ``X'' in Fig.~\ref{fig:principle}) may have been measured during the rise time of the peak detector, it does not accurately represent the initial pulse amplitude. Instead, we record the following five samples and use a simple exponential regression fit to calculate the amplitude at the time of the triggering sample. This is used as the measured peak amplitude.

As shown in Fig. \ref{fig:electronics}, the Arduino performs several other functions as well. It is used to:
\begin{enumerate}
\item update the OLED screen. The provided Arduino code (Arduino/Arduino\_code/Arduino\_code.ino) communicates with the OLED screen to output the number of observed events, run time, count rate, and a  bar indicating the pulse amplitude of the last event.
\item pulse an LED light with a pulse length proportional to the calculated SiPM pulse amplitude.
\item monitor the detector dead time. Each command issued to the Arduino increases the total amount of time in which the detector is unavailable to make a measurement. The Arduino code measures the time each command takes and subtracts it from the total detector live time. The main source of dead time at sea level is due to the time it takes to update the OLED screen. Each update, which happens every second, takes approximately 50~ms. The next largest source of dead time is flashing the LED proportionally to the number of photons the SiPM observed. The serial readout of the detector takes approximately 5~ms on average and is therefore not a significant component of the dead time.
\item communicate with a computer via the mini-USB terminal. This is used to record data directly to a computer through a serial port. The first three lines of the output are header file information. The remainder of the file saves data in the following format:
date stamp of the event given by Python, time stamp of the event given by Python, event number, time stamp of the event from the Arduino in milliseconds, measured pulse amplitude from the peak detector in volts, calculated SiPM pulse amplitude in mV, and measured dead time for a given event number in milliseconds.
\end{enumerate}

The code can be modified by installing the Arduino IDE and the required drivers for the Arduino Nano. To record data to the computer, one needs to ensure the Arduino Nano driver is properly installed and then run the Import\_data.py python program in the supplementary material  (Arduino/Import\_data.py). The python program will list the available serial ports for you to select from.

\subsection{Detector calibration}\label{sec:cal}

To determine how the measured pulse amplitude, given by the Arduino, corresponds to pulse amplitude from the SiPM, we remove the SiPM PCB and injected waveforms (of the same shape as SiPM pulses) of known amplitude into position A shown in Fig. \ref{fig:electronics}. The waveforms were generated by first measuring the amplitude of a SiPM pulse as a function of time, then inputting this information into a pulse generator. The pulse generator allowed us to scale the waveform to an arbitrary amplitude between 0 to 5000~mV. The waveforms were then injected into the circuit at a desired frequency.  The Arduino was then used to record the measured pulse amplitude after the peak detector circuit. With this, we are able to convert between Arduino measurements and SiPM outputs.  Fig.~15 shows the resulting measurements from input pulse amplitudes varying from 0 to 1000~mV. The pulses were injected into the circuit at a frequency of 20 Hz. There is a strong correlation between the input pulse amplitude and the measured pulse amplitude (light blue circles). A 2$^{\mathrm{nd}}$-order polynomial fit from 30 and 700~mV yields a relationship of:

\begin{equation} \label{eq:linear}
\mathrm{y}= -8.5432\times10^{-4}~\mathrm{x}^2~+ 1.7859 ~x~- 33.3687
\end{equation}
where y is the measured pulse amplitude from 0 to 1024 (0--5~V on the right axis) and x is the input pulse amplitude.  With the positive root of the quadratic equation, we can use the inverse of this equation to convert a measured pulse amplitude by the Arduino to an input pulse (SiPM) amplitude. 

The standard deviation of the 250 measurements for a given input voltage was found to be approximately 3.2~\%.

\begin{figure*}
\includegraphics[width=0.8\columnwidth]{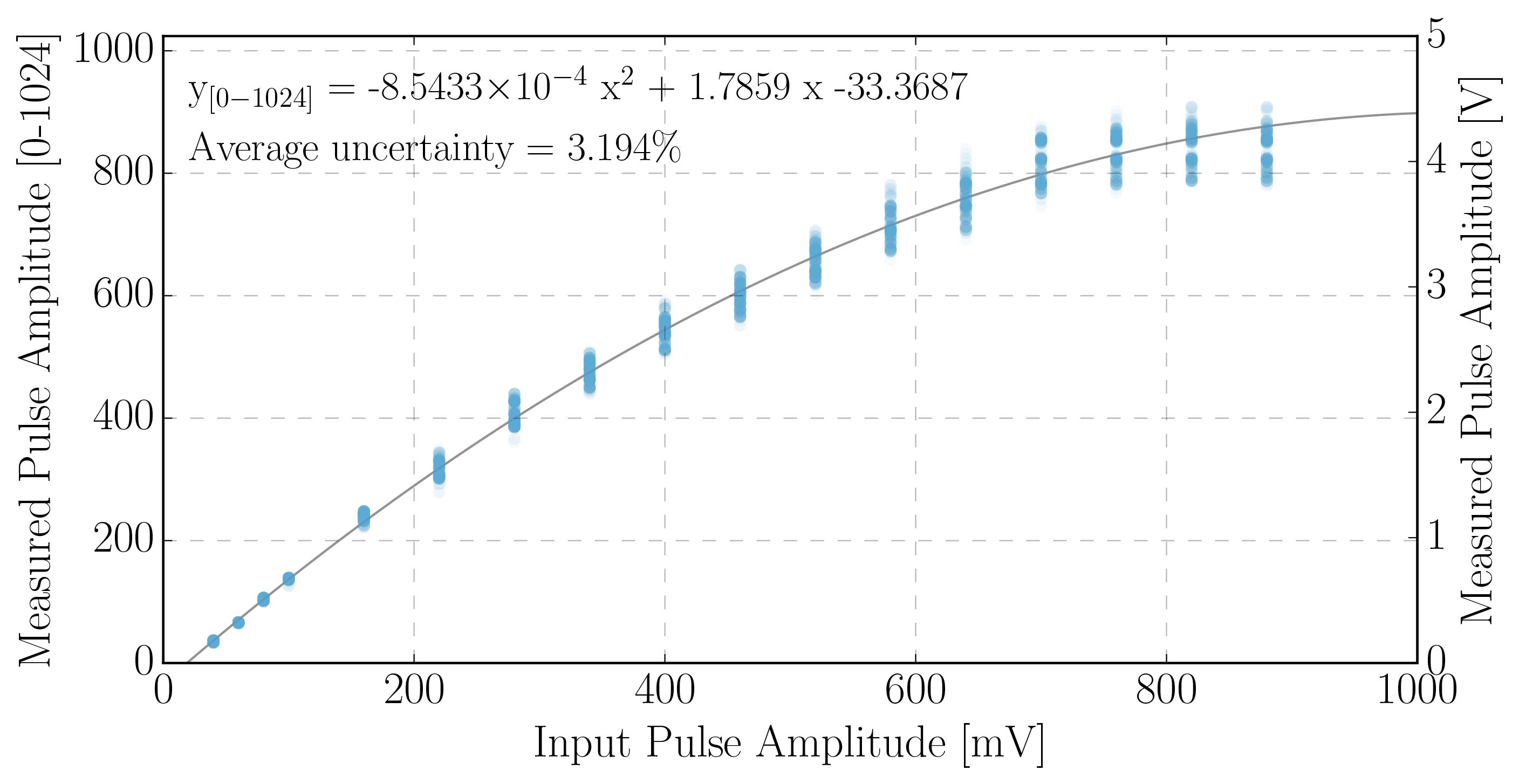}\label{fig:dat}
\caption{Calibration data for the complete circuit. There are 250 samples per input pulse amplitude (light blue circles). The measurements are semi-opaque to show the relative distribution. }
\end{figure*}

The data in Fig.~15 also shows that for input pulses with an amplitude greater than roughly 700~mV, the measured pulse amplitude becomes saturated at approximately 4.3~V. This is due to the limited voltage range on the op amp combined with the voltage drop across the diode in the peak detector circuit. While pulses with initial amplitudes greater than 700~mV are observed, they are relatively rare. 

\section{Learn about Cosmic Ray Muons}\label{sec:cr}

The desktop muon counter triggers on muons that are produced when high-energy astrophysical particles, called cosmic rays, collide with the Earth's atmosphere, producing particles that decay to muons. 
In his 1950 Nobel Lecture, C.F. Powell described cosmic rays as a ``thin rain of charged particles''~\cite{Powell}.
Most cosmic rays are produced in our galaxy and are nuclei expelled in supernova explosions.   About 90\% of cosmic rays are protons,  9\% are helium nuclei, and the remaining 1\% are heavier nuclei.  When cosmic rays hit the nuclei of the atmosphere, a shower of particles is produced, including pions and kaons.  These are the progenitors of the muons.  Students may be assigned to read three classic works by physicist Bruno Rossi about cosmic rays~\cite{Bruno1, Bruno2, Bruno3}.  The origin and content of cosmic rays remains a hot topic of study today, with major conferences devoted to the latest results~\cite{ICRC}.  A useful resource for lectures on cosmic rays is Chapter 28 of Ref.~\cite{PDG}, The Particle Data Book.  This summarizes our most up-to-date knowledge.

The muons that are ultimately produced in the shower are fundamental particles that carry electric charge of $+1$ or $-1$ and have mass that is about 200 times that of the electron.   For a brief introduction to muons and their place within the Standard Model of particle physics, we recommend that students visit The Particle Adventure website~\cite{ParticleAdventure}.  Muons are unstable and will decay to an electron, a neutrino, and an antineutrino.   At rest, the lifetime of the muon is approximately 2.2 microseconds.   Given that muons are produced in the shower at more than 10 km above the Earth's surface,  Galilean relativity calculations show a very small probability of survival to reach the desktop muon counter. However, because muons are produced at high energies, relativistic time dilation extends their lifetime.  As a result, muons can survive and be detected on Earth.  Calculation of the different expectations for Galilean and special relativity is a useful exercise for the student.

The muon flux at sea level is about one per square centimeter per minute for a horizontal detector~\cite{PDG}.
This constant bombardment by muons has pros and cons for a particle physicist. 
On the plus side, cosmic ray muons are commonly used in surface-based particle physics experiments in order to commission and calibrate detectors before they are exposed to beam produced by accelerators.    
Often the muons detected at sea level are accompanied by other particle debris, such a photons and protons.  A relatively small amount of shielding material is often used to remove this accompanying debris, leaving only the muons for use in calibration.  On the other hand, many particle physics experiments are looking for rare events, and the rare signal can be swamped by the muon signal.  These experiments must be located in deep underground laboratories. The U.S. is in the process of building a new deep underground laboratory in Lead, South Dakota, which is described in Ref.~\cite{SanfordLab}.

\section{Example measurements and final remarks}

To illustrate some of the capabilities of the detector and to hopefully inspire students to make their own measurements, we have used the desktop muon counter to make several measurements. This section includes a coincidence measurement between two detectors arranged to measure the angular distribution of cosmic ray muons and several rate measurements at various altitudes and levels of overburden. 

A common measurement to make with muon detectors is to determine the muon rate as a function of polar angle. According to the Particle Data Group, the angular muon dependence is proportional to cos$^2\,\theta$, where $\theta$ represents the polar angle with respect to vertical, for minimum ionizing muons with roughly 3~GeV of energy. For lower energy muons, the distribution becomes increasingly steep, while at much higher energies, it flattens out, approaching sec\,$\theta$ for directions near $\theta < 70^\circ$~\cite{PDG}. 

\begin{figure}[h!]
\centering
\includegraphics[width=0.6\columnwidth]{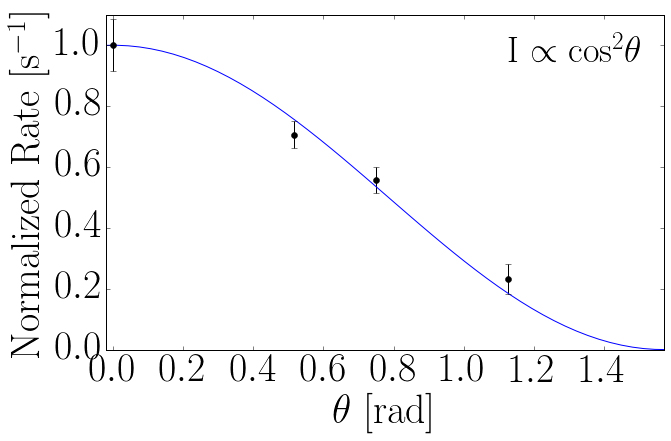}
    \caption{The relative rate as a function of polar angle. The data is shown in black and the theoretical minimum ionizing muon distribution is show in solid blue. }
    \label{fig:coincidence}
\end{figure}

We placed two detectors side by side in order to minimize the angular acceptance between the two pieces of scintillator. The data for both detectors was recorded on the same computer and given an uncertainty of the time stamp (given by the computer) of 5~ms to account for the serial communication delay. Data was recorded at several polar angles over the course of a day. The relative rate for the measurements is shown in Fig. \ref{fig:coincidence} and was found to be in good agreement with a cos$^2\,\theta$ dependence.

We also performed several measurements at high altitudes and at underground facilities with a significant amount of overburden. Overburden is shielding by overhead material that attenuates cosmic ray muons. Since the density of the material shielding from cosmic rays may vary, we define overburden in terms of the number of meters of water that would provide the same attenuation, abbreviated as meter-water-equivalent (m.w.e.).

\begin{figure}[h!]
\centering
\includegraphics[width=0.7\columnwidth]{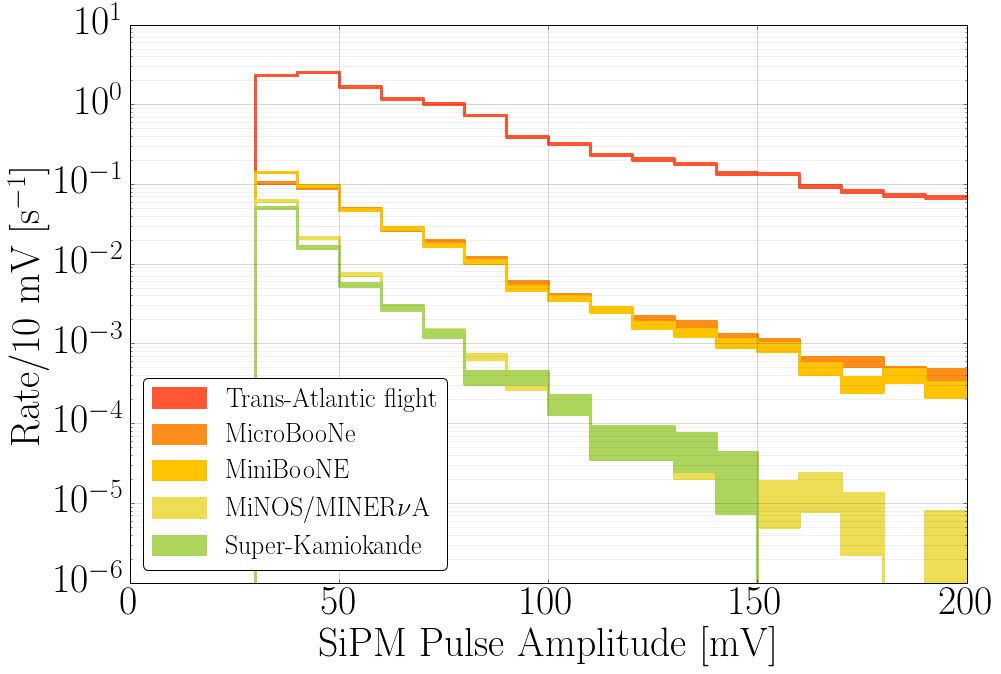}
    \caption{Sample data from measurements made at various overburdens and altitudes. We see several orders of magnitude change in the raw detector count rate between an airplane flight at 41,000~ft and underground at the Super-Kamiokande detector (2700~m.w.e. overburden).}
    \label{fig:measurements}
\end{figure}

\begin{figure}[h!]
\centering
\includegraphics[width=0.7\columnwidth]{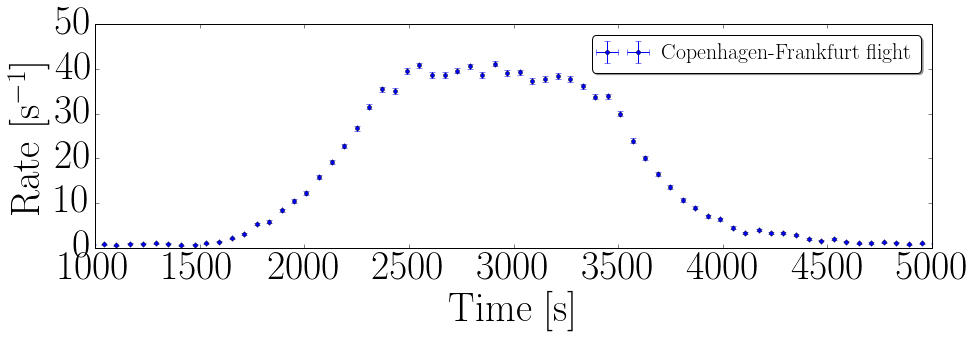}
    \caption{A rate measurement during a short flight. The maximum altitude was 33,000~ft.}
    \label{fig:flight}
\end{figure}

Fermilab is home to several high profile neutrino experiments, each of which utilizes different methods to remove the cosmic ray muon contamination. MINER$\nu$A~\cite{minerva} and the near detector of MINOS~\cite{minos}, for example, are buried over 100~m underground in order to attenuate the cosmic ray muon rate, whereas MicroBooNE~\cite{microboone} is located in a building with very little overburden. The MiniBooNE~\cite{miniboone} detector, on the other hand, was buried under several meters of soil inside of a concrete building. Rate measurements with the desktop muon detector were made at these locations over the period of a few days. The rates as a function of measured SiPM pulse amplitude are shown in Fig.~\ref{fig:measurements} along with a high-altitude measurement made during a Trans-Atlantic flight. 

Super-Kamiokande (SK)~\cite{sk} is a 50~kton Cherenkov neutrino detector located in the Kamioka mine, with a meter-water-equivalent  of overburden of approximately 2700~m. The measurement at SK was performed using a different detector, and therefore the SiPM pulse amplitude cannot be directly compared to the other measurements. However, the SK measurement does represent a signal that originates purely from background contamination. The detector is sensitive to $\alpha$, $\beta$, and $\gamma$ radiation; however, the aluminum enclosure and the foil surrounding the plastic scintillator is sufficiently thick to attenuate most $\alpha$ and $\beta$ radiation.

For a high-altitude rate measurement, we were given permission to record data in an airplane at 41,000~ft. We see roughly a 50x increase in rate compare to the ground-level measurement. This is near the peak in the cosmic ray muon production region at approximately 45,000--60,000~ft.  Another measurement was made during take-off and landing for a short flight. The resulting measurement is shown in Fig.~\ref{fig:flight}. It shows that we can easily identify the altitude of the airplane and correlate it to the cosmic ray muon flux. 

There are many interesting physics measurements that the desktop muon detector, alone or as an array of multiple detectors, can be used to measure. Variations of the project in the previous section could include:
\begin{enumerate}
\item Expand on the coincidence measurement presented at the beginning of this section by following the measurement outlined in \cite{angle}. They include a calculation on the finite size of a detector  and provide a more in-depth description of the measurement procedure. 
\item Add bluetooth, wifi, temperature sensors, or in-situ data storage with a microSD card reader to the Arduino to expand the capabilities of the detector. There is a large online community of Arduino users, and they have built up a pool of examples of how to implement these technologies.
\item Measure the relative depths of subway stations across the city using the measured muon rates. 
\item Test relativistic time dilation on the cosmic ray flux by measuring the flux at various elevations, such as in an airplane or on a mountain, compared to sea level~\cite{muon}.
\item Investigate seasonal variations in muon rates. The National Oceanic and Atmospheric Administration's National Weather Service \cite{noaa} records local atmospheric conditions that can be used to investigate weather and rate correlations. 
\item Determine the correlation between muon rate and altitude. This may require elevation differences of at least several hundreds of meters. Multiple detectors can be used in coincidence to improve the muon purity of the measurement. 
\item Use GEANT4 to simulate the angular response function and correlate the pulse height to the energy deposited in the scintillator. This requires the knowledge of the scintillator material and familiarity with C++.
\end{enumerate}

The construction of desktop muon detectors will teach useful skills in machine- and electronics-shop activities. The code, libraries, and technical drawings are all provided.  The time scale for a student to produce a muon detector is expected to be less than 100 hours. Once proficient with the machinery, we have found a student can produce approximately one detector per day. The total cost of a single detector is approximately \$100 and may decrease in time as SiPMs become less expensive. 

\begin{acknowledgments}
This work is supported by the NSF grant 1505858. The authors would like to thank SensL and Fermilab, for donations that made the development of this project possible, as well as P. Fisher at MIT and IceCube collaborators at WIPAC, for their support in developing this as a high school and undergraduate project. We also extend thanks to K. Frankiewicz for performing the measurements at Super-Kamiokande; J.~Moon, D.~Torretta, and J.~Zalesak for their aid in making the measurements at Fermilab;  B. Jones for the idea of developing this project for the IceCube detector; and those who taught Phys 063 at Swarthmore College for the inspiration.

\end{acknowledgments}

\pagebreak

\end{document}